\documentclass[pra,aps,twocolumn,superscriptaddress,tightenlines]{revtex4-2}
\usepackage{graphicx}
\usepackage{bm}
\usepackage{braket}
\usepackage{pifont}
\usepackage{amsmath}
\usepackage{amsfonts}
\usepackage{amssymb}    
\usepackage{color,xcolor}
\usepackage[caption=false]{subfig}
\usepackage{natbib}
\usepackage{dsfont}
\usepackage{hyperref}
\hypersetup{
	colorlinks=true,
	linkcolor=blue,
	filecolor=black,  
	citecolor=blue, 
	urlcolor=blue,
	pdftitle={},
	bookmarks=true,
}

\begin{document}
\title{Continuous variable quantum key distribution through a fading channel with minimal information leakage}
\author{Josephine Dias} 
\email{j.dias@uq.edu.au}
\affiliation{Centre for Quantum Computation and Communication Technology, School of Mathematics and Physics, University of Queensland, St Lucia, Queensland 4072, Australia}
\author{Timothy C. Ralph} 
\affiliation{Centre for Quantum Computation and Communication Technology, School of Mathematics and Physics, University of Queensland, St Lucia, Queensland 4072, Australia}
\date{\today}
\begin{abstract}
Continuous variable quantum key distribution enables two parties to extract provably secure keys for cryptography by exploiting aspects of quantum mechanics. A central challenge in achieving QKD over long distances is overcoming optical loss, which limits the secret key rates and achievable transmission distances. Recent work has illustrated how to perform continuous variable quantum key distribution while eliminating information leakage to an eavesdropper in a pure-loss channel and minimizing information leakage in a noisy channel. In this work, we analyse this protocol under conditions where the quantum channel linking the sender and receiver is a fading channel, which can be used to model free-space links. We find that the protocol maintains strong performance over fading channels and outperforms comparable protocols.
\end{abstract}
\maketitle

\section{Introduction}
One of the most mature applications of quantum technology can be found in quantum key distribution (QKD) where Heisenberg's uncertainty principle is exploited in order to distribute a key for cryptography with information theoretic security. This is in contrast to current cryptographic  protocols which rely on computational problems being hard to solve. QKD was first conceived in a discrete variable approach, where the key was encoded in two different non-orthogonal bases of a two-level system \cite{bennett1984quantum, ekert1991quantum}. Alternative approaches using continuous variables (CV) soon appeared \cite{ralph2000security, grosshans2002continuous, laudenbach2018continuous, pirandola2020advances, usenko2026continuous}. The promise of higher key rates as well as better compatibility with existing infrastructure makes continuous variable quantum key distribution an attractive implementation.

To perform quantum key distribution between two parties, a quantum channel, through which signals are sent between the two parties, is needed. This could be an optical fiber channel or a free space channel. Optical fibre links form the backbone of modern communication networks, enabling stable, high-capacity, low-loss transmission over long distances.  Free-space communication channels are of interest because they enable flexible links, including satellite and ground-based connections, where deploying optical fibre is impractical or impossible. While the transmissivity of an optical fibre link is generally stable, the transmissivity of a free-space channel fluctuates in time because of atmospheric, weather, and beam wandering effects producing an effect known as channel fading. Fading noise places a strong restriction on the range of CVQKD in free space channels  \cite{usenko2012entanglement, peuntinger2014distribution, hosseinidehaj2015gaussian, hosseinidehaj2021composable}. 

In a quantum key distribution protocol, Alice and Bob exchange signals while an eavesdropper, usually denoted Eve can interact with and make measurements on the exchanged signals. Security proofs for QKD consider a hierarchy of attack models including individual, collective and coherent, with coherent attacks being the most powerful type of attack. In an individual attack, Eve interacts with each transmitted signal separately and measures her ancilla. In a collective attack, Eve interacts each transmitted quantum state with an identically prepared ancillary system, stores her ancillas in a quantum memory, and performs a joint measurement after Alice and Bob reveal their classical information.
 
Recent results have investigated CV-QKD protocols in which the transmitted states are engineered to suppress the information leakage to an eavesdropper. In Ref.~\cite{jacobsen2018complete}, a restricted Gaussian alphabet of squeezed states was shown theoretically and experimentally to eliminate Eve's Holevo information in a purely lossy channel and to reduce it in the presence of channel noise. Ref.~\cite{winnel2021minimization} extended the security analysis to general asymmetric collective attacks and proposed a heralded symmetrization procedure to remove the security complications arising from the quadrature asymmetry. In Ref.~\cite{hosseinidehaj2022simple}, a related protocol based on a squeezed and amplitude-modulated laser was analysed in the composable finite-size regime for a free-space fading channel, under the assumption that classical monitoring restricts Eve to passive attacks. In contrast, here we consider active eavesdropping and analyse the protocol under both finite size individual and collective attacks and fading channels.
\begin{figure}
	\centering
	\includegraphics[width=0.98\linewidth]{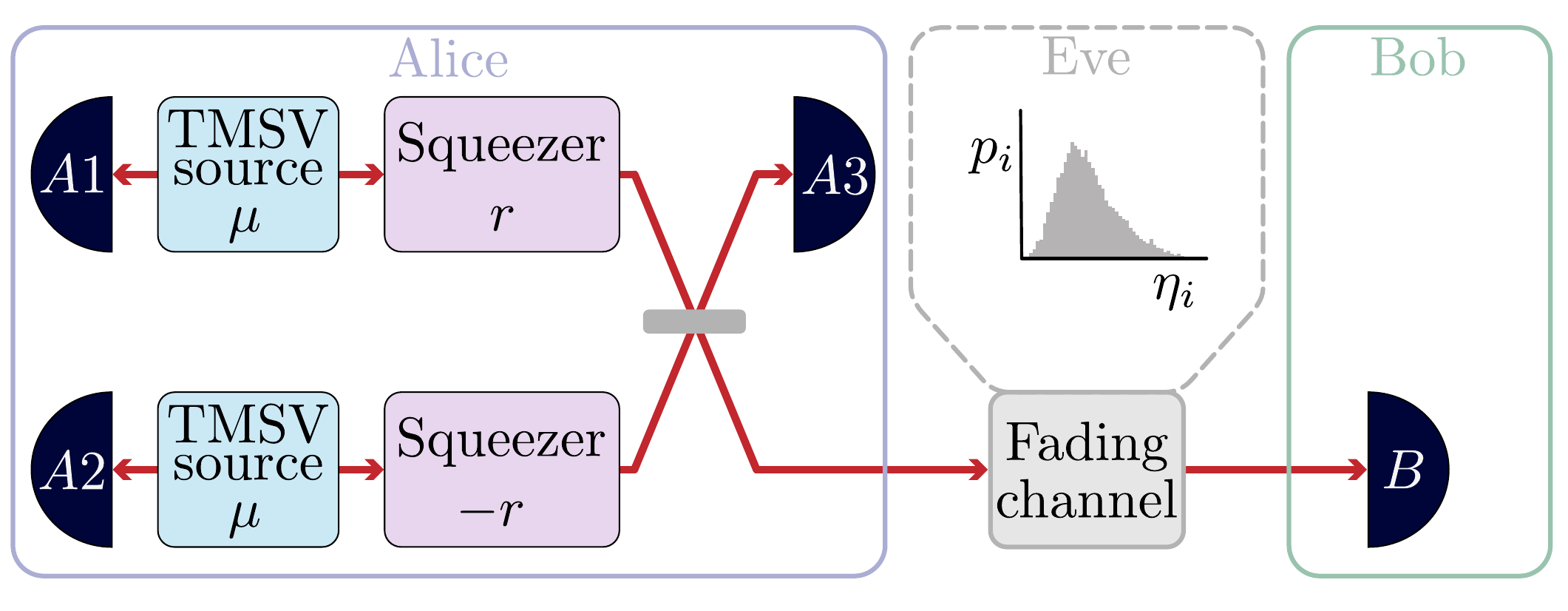}
	\caption{ Protocol for CVQKD with minimisation of information leakage to Eve.  The box labelled Alice contains the necessary operations for Alice to perform in order to minimise information leakage to Eve \cite{winnel2021minimization}. Squeezing is applied to one mode of two TMSV states which are then combined and mode A3 is input into homodyne detection. The remaining mode passes through a fading channel which has transmission that varies in time before arriving at Bob's station. }
	\label{fig:diagram}
\end{figure}

In this paper, we will show that the heralded scheme from Winnel et al. \cite{winnel2021minimization} which achieves elimination of information leakage to Eve in the pure-loss case and minimisation of information leakage to Eve in the noisy channel case has significant promise for use in free-space fading channels against active attacks. This paper is arranged in the following way: in Sec.~\ref{sec:background} we cover the necessary background for continuous variable quantum key distribution including how to calculate key rates, different types of attacks and different types of channels. In Sec.~\ref{sec:results} we explain our results, including that under the assumption of individual attacks this protocol can far outperform other comparable protocols. 

\section{Background \label{sec:background}}
\subsection{Continuous variable QKD}
In the asymptotic regime, the achievable CV-QKD key rate can be lower bounded by comparing the correlations shared by Alice and Bob with the maximum information accessible to Eve.  For collective attacks, this is bounded by the Devetak-Winter rate representing the minimum secret key rate achievable by Alice and Bob\cite{devetak2005distillation}:
\begin{equation}
K=\beta I_{AB}-\chi_{EB}
\label{eq:KR}
\end{equation}
where $I_{AB}$ is the mutual information between Alice and Bob and $\chi_{EB}$ is the Holevo bound, the upper limit on information that Eve can extract from Alice and Bob's quantum system. The parameter $\beta$ captures reconciliation efficiency which may be less than 1. Note that the Devetak-Winter rate is an asymptotic result, meaning it captures the rate achievable when Alice and Bob can exchange an infinitely large number of signals ($N\to \infty$). However, in real-world scenarios the finite sized number of signals that Alice and Bob can exchange will affect the number of secure secret key bits. In practice, these finite-size effects cause the estimation of the channel parameters, transmission and excess noise, to have some uncertainty on their estimations. As a consequence, Eve's information cannot be known with absolute certainty. 

Estimating the amount of information that Eve can ascertain about the ensemble shared between Alice and Bob necessarily means making assumptions on the capabilities that Eve is able to achieve. For collective attacks, Eve's accessible information is bounded by the Holevo quantity,
\begin{equation}
	\chi_{EB}
	=
	S(\rho_E)-S(\rho_{E|b}).
\end{equation}
By assigning Eve the purification of the state $\rho_{AB}$ shared by Alice and Bob, the entropy identities $S(\rho_E)=S(\rho_{AB})$ and $S(\rho_{E|b})=S(\rho_{A|b})$ may be used, giving
\begin{equation}
	\chi_{EB}
	=
	S(\rho_{AB})-S(\rho_{A|b}),
\end{equation}
where $\rho_{A|b}$ denotes Alice's state conditioned on Bob's measurement outcome $b$. For Gaussian states, these entropies are calculated from the symplectic eigenvalues $\nu_k$ of the corresponding covariance matrix using $S(\rho)=\sum_{k=1}^{N}\left[\frac{\nu_k+1}{2}\log_2\left(\frac{\nu_k+1}{2}\right)-\frac{\nu_k-1}{2}\log_2\left(\frac{\nu_k-1}{2}\right)\right]$, where $N$ is the number of modes.

In an individual entangling-cloner attack, Eve replaces the channel use with a beam splitter reproducing the effective transmission of the channel, and mixes Alice’s signal with one mode of an EPR state whose variance reproduces the channel excess noise, and independently measures the modes retained from each transmission. For individual attacks, $\chi_{EB}$ in \eqref{eq:KR} must be replaced by $I_{BE}$. We outline how $I_{EB}$ is calculated in the Appendix (Sec.~\ref{sec:IBE}). 
\subsection{Heralding protocol}
We consider the heralded prepare-and-measure protocol introduced in Ref.~\cite{winnel2021minimization}. Alice initially generates two independent ensembles of squeezed states. A Gaussian displacement is applied to only one quadrature of each ensemble: the first is modulated in $x$, while the second is modulated in $p$. We denote the squeezed-quadrature variance by $V_{\mathrm{S}}$ and the modulation variance by $V_{\mathrm{M}}$, with the shot-noise variance normalised to unity. The two modes are then interfered on a balanced beam splitter. Alice performs homodyne detection on one output mode, labelled $A3$, randomly selecting either the $x$ or $p$ quadrature, while the other output mode, labelled $B$, is transmitted to Bob. Bob independently homodynes either $x$ or $p$. During classical post-processing, Alice and Bob announce their measurement bases and retain only those events for which they measured the same quadrature. 

For an ideal pure-loss channel, the information available to an eavesdropper can be eliminated when the variance of the relevant quadrature of the heralded state is equal to the shot-noise variance. In terms of the parameters defined above, this condition is
\begin{equation}
	V_{\mathrm{M}}
	=
	\frac{V_{\mathrm{S}}^{2}-2V_{\mathrm{S}}+1}
	{2-V_{\mathrm{S}}}.
\end{equation}
Thus, for a chosen squeezing variance $V_{\mathrm{S}}$, the modulation variance $V_{\mathrm{M}}$ is fixed by this relation. This zero-leakage condition applies to the ideal pure-loss channel; fading and excess noise generally prevent the complete elimination of Eve's information.

In the entanglement-based (EB) implementation (pictured in Fig.~\ref{fig:diagram}), Alice's state preparation is replaced by Einstein--Podolsky--Rosen (EPR) states together with squeezing applied to one mode of each pair. Specifically, mode $B$ is squeezed in the $x$ quadrature with squeezing parameter $r$, while mode $A3$ is squeezed by the same amount in the $p$ quadrature. The EB implementation is equivalent to the prepare-and-measure (PM) protocol when Alice performs homodyne detection on mode $A1$ in the $x$ quadrature and mode $A2$ in the $p$ quadrature.

For the heralded protocol of Ref.~\cite{winnel2021minimization}, the
secret key is extracted from the $x$-quadrature measurements of modes
$A_1$ and $B$, conditioned on a successful $x$-quadrature homodyne
measurement of mode $A_3$. The mutual information between Alice and Bob
is calculated as
\begin{equation}
	I_{AB}
	=
	\frac{1}{2}
	\log_2
	\left(
	\frac{V_{B_x}}
	{V_{B_x\mid A_{1,x}}}
	\right),
	\label{eq:IAB}
\end{equation}
where $V_{B_x}$ is Bob's unconditional $x$-quadrature variance and $V_{B_x\mid A_{1,x}}$ is the variance remaining after conditioning on Alice's $x$-quadrature measurement of mode $A_1$.

\subsection{Fading channel}
We model the fading channel as an ensemble of subchannels assuming the transmissivity in each subchannel is sufficiently stable. The $i$th subchannel has intensity transmissivity $\eta_i$ and occurs with probability $p_i$, where \begin{equation} \sum_i p_i=1. \end{equation} We assume that the excess noise $\xi$ is constant across all subchannels. The output of a fading channel is an ensemble average over the states produced by its constituent subchannels, with the corresponding Wigner function given by the probability-weighted average of their individual Wigner functions \cite{dong2010continuous}.

For a two-mode initial state, the covariance matrix of the
ensemble-averaged state after the fading channel can be reproduced,
at the level of its first and second moments, by an effective
fixed-loss channel. The effective transmissivity is
$T=\braket{\sqrt{\eta}}^2$. The covariance matrix after the fading
channel is
\begin{equation}
	\Gamma_{AB_1}^{\mathrm{fad}}
	=
	\left[
	\begin{array}{cccc}
		a_x  & 0    & c_x' & 0    \\[3pt]
		0    & a_p  & 0    & c_p' \\[3pt]
		c_x' & 0    & b_x' & 0    \\[3pt]
		0    & c_p' & 0    & b_p'
	\end{array}
	\right].
	\label{eq:cmFad}
\end{equation}
The correlations are
\begin{equation}
	c_x'=\sqrt{T}\,c_x,
	\qquad
	c_p'=\sqrt{T}\,c_p.
\end{equation}
The output variances are
\begin{align}
	b_x'
	&=
	Tb_x+1-T+T\xi
	+\operatorname{Var}\left(\sqrt{\eta}\right)(b_x-1),
	\label{eq:bxFad}
	\\
	b_p'
	&=
	Tb_p+1-T+T\xi
	+\operatorname{Var}\left(\sqrt{\eta}\right)(b_p-1).
\label{eq:bpFad}
\end{align}
Here, $\xi$ denotes the excess noise of the effective
fixed-transmissivity channel. The total effective excess noise,
including the contribution arising from fading, is defined separately
for each quadrature as
\begin{align}
	T\xi_{f,x}
	&=
	T\xi
	+\operatorname{Var}\left(\sqrt{\eta}\right)(b_x-1),
	\label{eq:Txix}
	\\
	T\xi_{f,p}
	&=
	T\xi
	+\operatorname{Var}\left(\sqrt{\eta}\right)(b_p-1).
	\label{eq:Txip}
\end{align}
The parameters describing the fading channel are
\begin{equation}
	T=\braket{\sqrt{\eta}}^2,
	\qquad
	\operatorname{Var}\left(\sqrt{\eta}\right)
	=
	\braket{\eta}
	-
	\braket{\sqrt{\eta}}^2.
\end{equation}

From \eqref{eq:Txix} - \eqref{eq:Txip}, it can be seen that  the fading-induced noise scales linearly with the variance of the amplitude transmissivity, $\operatorname{Var}(\sqrt{\eta})$, and with the input variance above shot noise. 

\section{Results \label{sec:results}}

\subsection{Asymptotic}
\begin{figure}
	\centering
	\includegraphics[width=0.98\linewidth]{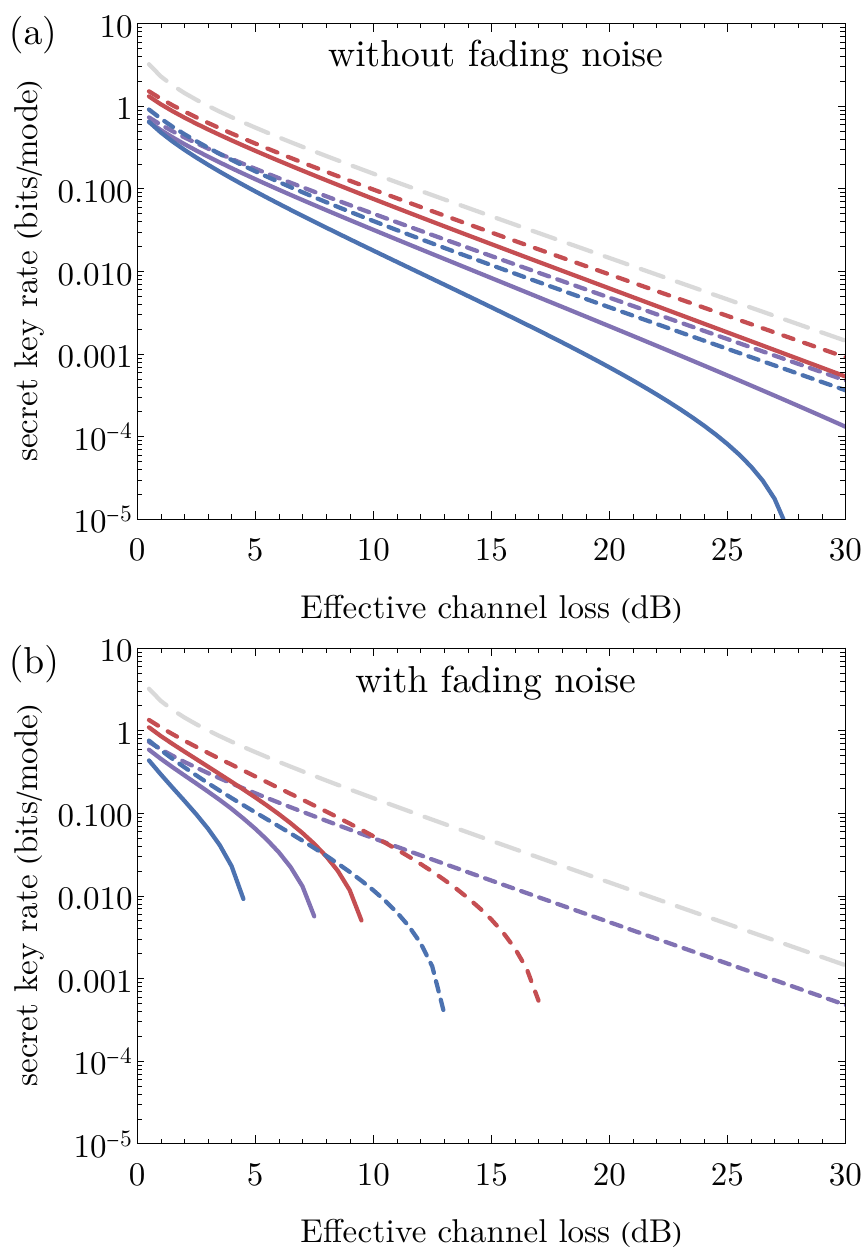}
	\caption{ Secret key rate of the heralding protocol (purple) against collective (solid) and individual (dashed) attacks in a fading channel. For comparison, we show the secret key rates of a squeezed state protocol with homodyne detection (red)\cite{cerf2001quantum} and a coherent state protocol with heterodyne detection (blue)\cite{weedbrook2004quantum}. All protocols have been limited to 10dB of squeezing.  The horizontal axis shows the effective channel loss,
		$L=-10\log_{10}T$, expressed in decibels, where  $T=\langle\sqrt{\eta}\rangle^{2}$ is the effective channel transmissivity. All protocols have been calculated using the fading channel model \eqref{eq:cmFad}-\eqref{eq:bpFad} with an excess noise of $\xi=0.05$. The gray, dotted line depicts the repeaterless PLOB bound (assuming no fading channel noise or excess noise) \cite{pirandola2017fundamental}. In (a), the amplitude-transmissivity variance is $\operatorname{Var}(\sqrt{\eta})=0$ and in (b) we use $\operatorname{Var}(\sqrt{\eta})=0.01$.} 
	\label{fig:KRA}
\end{figure}

In Fig.~\ref{fig:KRA}, we present the main asymptotic result of this paper comparing the secret key rates of the heralding protocol with those of two established Gaussian CVQKD protocols over a fading channel. The results for the heralding protocol are shown in purple, while the red and blue curves correspond, respectively, to a squeezed-state protocol with homodyne detection \cite{cerf2001quantum} and a coherent-state protocol with heterodyne detection \cite{weedbrook2004quantum}. Dashed curves represent individual attacks, whereas solid curves represent the stronger collective-attack model. All three protocols are restricted to a maximum squeezing of $10\,\mathrm{dB}$ and are evaluated using an excess noise of $\xi=0.05$. This figure illustrates the effect of fading noise on the key rates, with no fading noise in panel (a) and $\operatorname{Var}(\sqrt{\eta})=0.01$ in panel (b). We choose \(\operatorname{Var}(\sqrt{\eta})=0.01\) to represent strong fading, consistent with the magnitude of fluctuations inferred from numerical simulations of ground-to-satellite uplinks that account for diffraction, atmospheric extinction, turbulence and pointing errors \cite{zuo2023suppression}.

When Eve is restricted to individual measurements, the heralding protocol exhibits substantially greater robustness to channel loss. At low loss, the conventional squeezed-state protocol achieves a higher key rate; however, its key rate decreases rapidly and vanishes at approximately $17\,\mathrm{dB}$. Similarly, the coherent-state protocol ceases to generate a positive key at approximately $13\,\mathrm{dB}$. In contrast, the key rate of the heralding protocol remains positive throughout the displayed range, extending beyond $30\,\mathrm{dB}$ of effective channel loss. The purple dashed curve also follows a nearly constant slope on the logarithmic scale, indicating that the protocol strongly suppresses the information available to Eve under an individual attack, despite the presence of fading-induced noise.

The improvement is considerably reduced when Eve is permitted to perform collective measurements. In this case, the key rates of all three protocols fall sharply and vanish at much lower losses. Although the protocol is highly effective at limiting the information obtained from individual measurements, Eve's ability to retain her quantum states and perform a collective measurement seems to substantially increase the information available to her. For reference, the gray dashed curve shows the repeaterless PLOB bound for an ideal pure-loss channel. Since this bound is evaluated without fading or excess noise, it provides an idealised reference, not a direct comparison under the same channel conditions.

\subsection{Suppression of fading-induced noise}
\begin{figure}
	\centering
	\subfloat[]{%
		\label{fig:eveInfo}
		\includegraphics[width=0.98\linewidth]{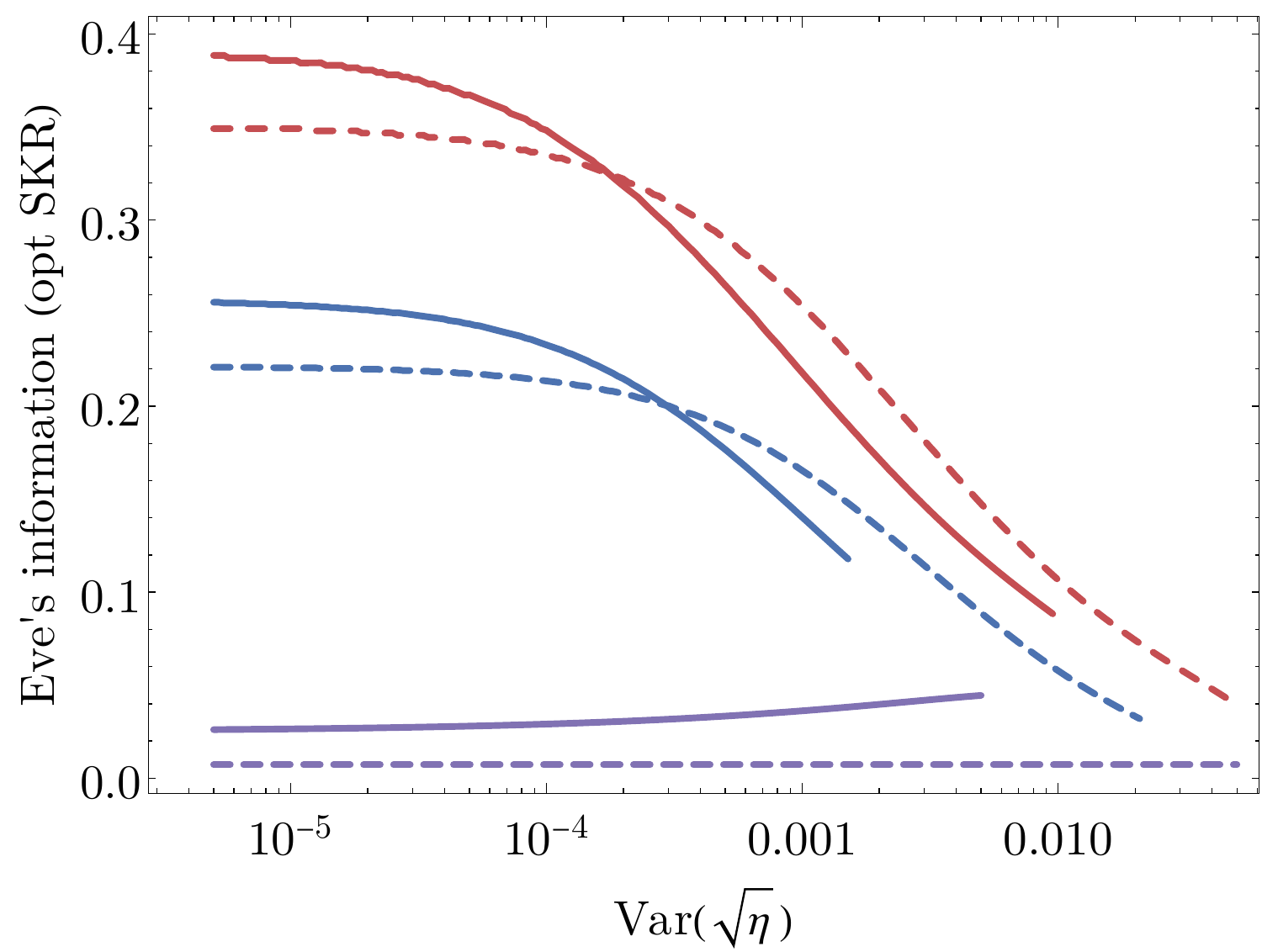}
	}%
	\hfill % Adds horizontal spacing between the two images
	\subfloat[]{%
		\label{fig:IAB}
		\includegraphics[width=0.98\linewidth]{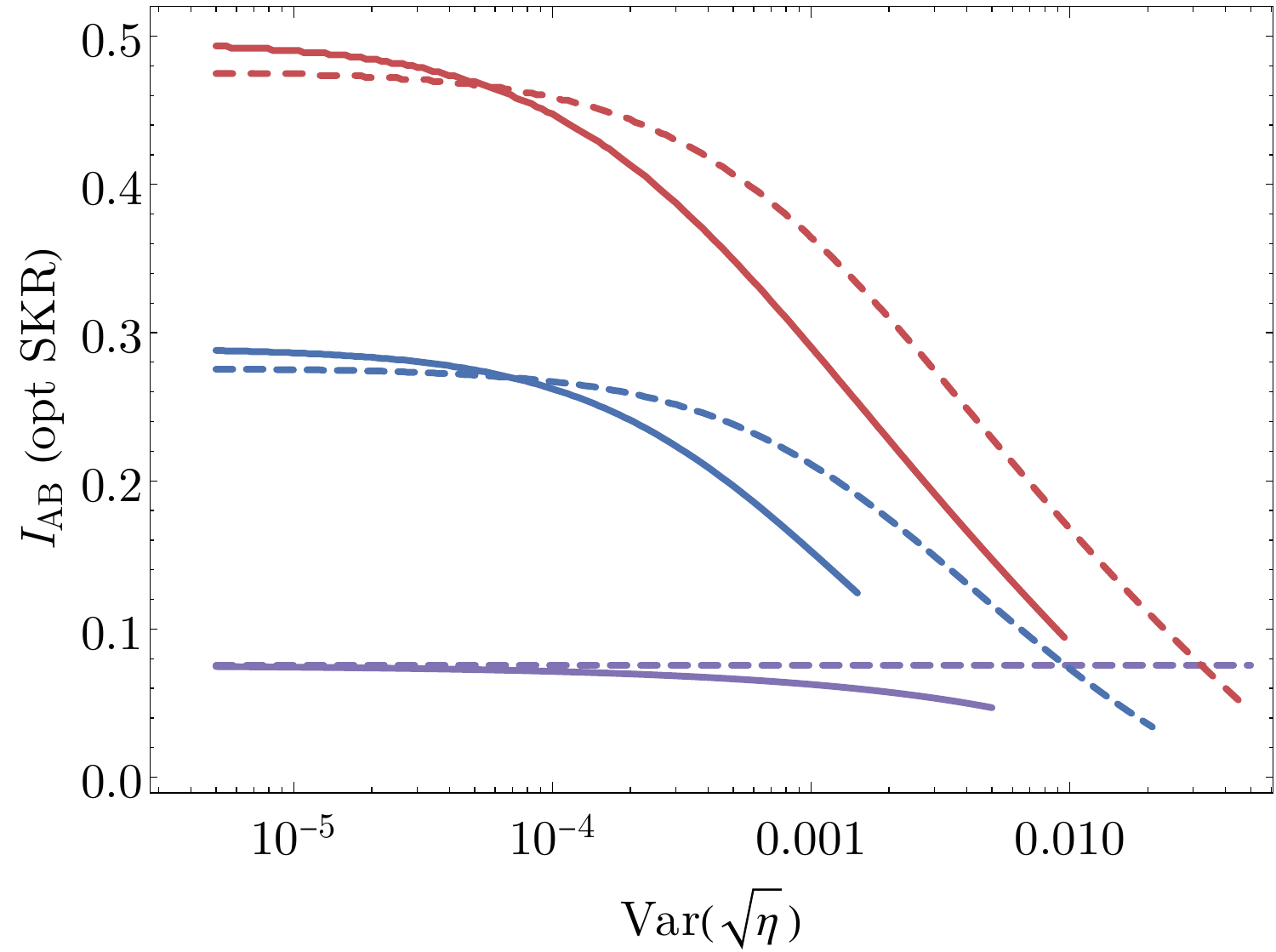}
	}
	\caption{Components in the optimised secret key rate for various fading noise parameters. In \protect\subref{fig:eveInfo}, we give Eve's information for increasing fading channel noise. In~\protect\subref{fig:IAB}, we give the mutual information between Alice and Bob for increasing fading noise. Line styles and colours are same as Fig.~\ref{fig:KRA}. In this figure we have set effective transmission to $T=0.1$, excess noise $\xi=0.05$ and optimised secret key rate (opt SKR) over all possible squeezing. We can see that increasing the fading noise affects all schemes except the heralding scheme under individual attacks. Lines stop at the point where no positive key can be achieved.}
	\label{fig:eveIAB}
\end{figure}
The markedly different behaviour of the heralding protocol under individual and collective attacks can be understood by examining the contributions to the secret key rate. As noted in Ref.~\cite{hosseinidehaj2022simple}, the noise arising from fluctuations in the channel transmissivity vanishes when the variance of the quadrature transmitted through the channel is equal to the shot-noise variance. In the effective fading-channel model used here, Bob's output variance in the key-generating quadrature is
\begin{equation}
	b_x'
	=
	Tb_x+1-T+T\xi
	+
	\operatorname{Var}(\sqrt{\eta})(b_x-1),
\end{equation}
where the shot-noise variance is normalised to unity. The contribution arising specifically from fading is 
\begin{equation}
	\operatorname{Var}(\sqrt{\eta})(b_x-1),
\end{equation}
which satisfies
\begin{equation}
	\operatorname{Var}(\sqrt{\eta})(b_x-1)=0
	\qquad
	\text{when}
	\qquad
	b_x=1.
\end{equation}
This cancellation applies only to the fading-induced contribution and does not remove the fixed excess-noise term $T\xi$.

The heralding protocol is designed so that the variance of the key-generating quadrature can be fixed at the shot-noise level by choosing the squeezing and modulation parameters according to the zero-leakage condition. By contrast, the total modulated quadrature variances of the conventional squeezed- and coherent-state protocols generally differ from unity. Consequently, increasing $\operatorname{Var}(\sqrt{\eta})$ introduces additional variance-dependent noise into these protocols, reducing the correlations shared by Alice and Bob.

To illustrate this behaviour, Fig.~\ref{fig:eveIAB} shows the two components of the optimised secret key rate as functions of $\operatorname{Var}(\sqrt{\eta})$. Fig.~\ref{fig:eveIAB}(a) shows Eve's information, while Fig.~\ref{fig:eveIAB}~(b) shows the mutual information $I_{AB}$ between Alice and Bob. The colours and line styles follow those used in Fig.~\ref{fig:KRA}: purple, red and blue represent the heralding, squeezed-state and coherent-state protocols, respectively, while dashed and solid curves correspond to individual and collective attacks.

Under individual attacks, Eve's information for the heralding protocol remains small and is independent of the fading variance, as shown by the purple dashed curve in Fig.~\ref{fig:eveIAB}(a). At the same time, the corresponding Alice-Bob mutual information in Fig.~\ref{fig:eveIAB}(b) remains approximately constant. The difference between these two quantities therefore remains positive over a broad range of fading strengths, explaining the persistence of the purple dashed key-rate curve in Fig.~\ref{fig:KRA}. Although the heralding protocol produces a smaller value of $I_{AB}$ than the two benchmark protocols at weak fading, the information available to Eve is also substantially smaller.

For the squeezed- and coherent-state protocols, $I_{AB}$ decreases as the fading variance increases. Eve's information also decreases along the optimised curves; however, this does not indicate an improvement in security. Rather, the reduction in $I_{AB}$ is sufficiently strong that the difference between Alice and Bob's reconciled information and Eve's information eventually becomes non-positive. The curves terminate at the points beyond which a positive secret key rate cannot be obtained. For the heralding protocol, the cancellation of the fading-induced noise in the $x$ quadrature does not provide the same level of protection against collective attacks. In this case, Eve's information is bounded by the Holevo quantity, which depends on the full covariance matrix, including the conjugate quadrature and the correlations between the different modes. 

\subsection{Finite-size effects}
\begin{figure}
	\centering
	\includegraphics[width=0.98\linewidth]{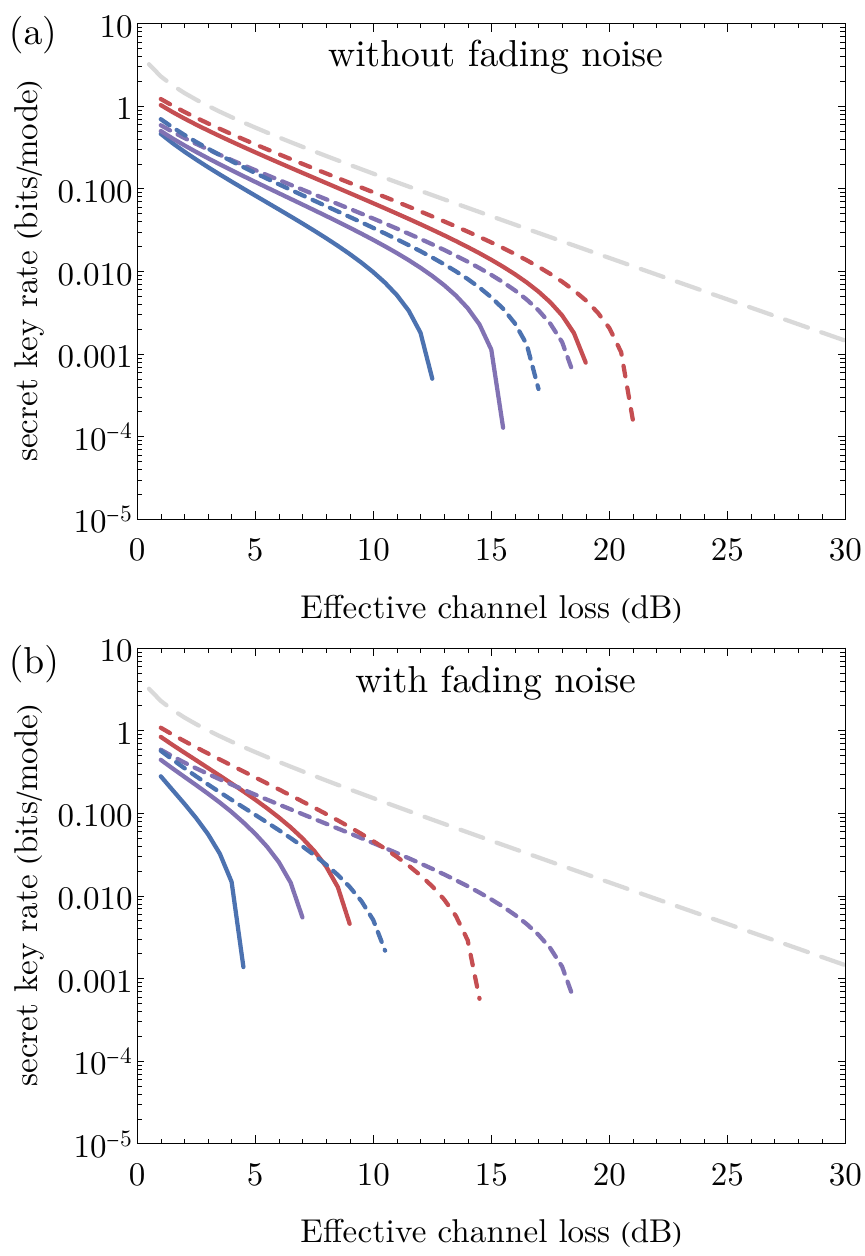}
	\caption{ Secret key rate of the heralding protocol (purple) against collective (solid) and individual (dashed) attacks in a fading channel. Here finite size effects are taken into consideration using a block size of $N=10^{10}$. For comparison, we show the secret key rates of a squeezed state protocol with homodyne detection (red) and a coherent state protocol with heterodyne detection (blue).  The gray, dotted line depicts the repeaterless bound \cite{pirandola2017fundamental}. In (a), the amplitude-transmissivity variance is $\operatorname{Var}(\sqrt{\eta})=0$ and in (b) we use $\operatorname{Var}(\sqrt{\eta})=0.01$.}
	\label{fig:KRF}
\end{figure}

In the previous section, we examined the performance of the heralding protocol in the asymptotic limit of infinitely many signals exchanged between Alice and Bob. In practice, only a finite number of signals is available, introducing uncertainty in parameter estimation and corrections to the achievable secret key rate. We now examine the performance of the protocol when these finite-size effects are included.

We adopt the finite-size treatment of Jain et al. \cite{jain2022practical}. For a key-generation block containing \(N\) modes before information reconciliation, the extracted key length is chosen according to
\[
\begin{aligned}
	\ell \leq{}&
	p_f N K_{\mathrm{PE}}^{\infty}
	-\sqrt{p_fN}\,\Delta_{\mathrm{AEP}}\\
	&-\sqrt{p_fN}\log_2(p_fN)\,\Delta_{\mathrm{ent}}
	+\Delta_S+\Delta_H ,
\end{aligned}
\]
where \(p_f=1-\mathrm{FER}\) is the probability of successful information reconciliation where $\mathrm{FER}$ is the frame error rate. The asymptotic contribution incorporates parameter-estimation uncertainty through
\[
K_{\mathrm{PE}}^{\infty}
=
\begin{cases}
	\beta I_{AB}-I_{EB}^{\mathrm{PE}},
	&\text{individual attacks},\\[2pt]
	\beta I_{AB}-\chi_{EB}^{\mathrm{PE}},
	&\text{collective attacks}.
\end{cases}
\]
We evaluate the following lower bound on the achievable finite-size
secret key rate, expressed in bits per mode:
\begin{equation}
	K_F
	=
	p_f K_{\mathrm{PE}}^\infty
	-
	\sqrt{\frac{p_f}{N}}
	\left[
	\Delta_{\mathrm{AEP}}
	+
	\log_2(p_f N)\,\Delta_{\mathrm{ent}}
	\right]
	+
	\frac{\Delta_S+\Delta_H}{N},
	\label{eq:finite_key_rate}
\end{equation}
where \(N\) denotes the block size.

The correction terms are
\[
\Delta_{\mathrm{AEP}}
=
4(d_{\mathrm{rx}}+1)
\sqrt{
	\log_2\!\left[
	\frac{2}{\left(p_f\varepsilon_s^2/3\right)^2}
	\right]},
\]
\[
\Delta_{\mathrm{ent}}
=
\sqrt{2\log_2\!\left(\frac{2}{\varepsilon_{\mathrm{ent}}}\right)},
\]
\[
\Delta_S
=
\log_2\!\left(p_f-\frac{p_f\varepsilon_s^2}{3}\right),
\]
and
\[
\Delta_H
=
2\log_2\!\left(\sqrt{2}\,\varepsilon_h\right).
\]
Here, \(d_{\mathrm{rx}}\) is the digitisation depth per quadrature. The parameters \(\varepsilon_s\), \(\varepsilon_{\mathrm{ent}}\) and \(\varepsilon_h\) specify the smoothing, entropy-estimation and hashing tolerances, respectively. The terms account for the finite-block entropy penalty, uncertainty in the empirical entropy, conditioning on successful reconciliation, and privacy amplification (see Appendix for details).

For collective attacks, \(\chi_{EB}^{\mathrm{PE}}\) is evaluated using conservative covariance-matrix estimates obtained from the parameter-estimation confidence intervals \cite{jain2022practical}. For individual attacks, \(I_{EB}^{\mathrm{PE}}\) is evaluated over the corresponding confidence region of the effective channel parameters \cite{hosseinidehaj2021composable}. 

The resulting key rates are presented in Fig. 4, where panel (a) shows the absence of fading noise and panel (b) includes fading characterised by \(\operatorname{Var}(\sqrt{\eta})=0.01\). Without fading, the squeezed-state protocol with homodyne detection achieves the highest rates over most of the plotted range. Against collective attacks, it also tolerates the greatest channel loss, while the heralding protocol extends the loss range beyond that of the coherent-state protocol with heterodyne detection. 

Fading substantially reduces the loss range of all three protocols against collective attacks. The heralding protocol continues to outperform the coherent-state benchmark at higher losses, although the squeezed-state protocol remains the most tolerant to loss. Against individual attacks, the heralding curve remains the same as its counterpart without fading, extending to approximately \(18\,\mathrm{dB}\), compared with approximately \(14\,\mathrm{dB}\) for the squeezed-state protocol and \(10\,\mathrm{dB}\) for the coherent-state protocol. The advantage of heralding in suppressing the effect of fading therefore remains evident when finite-size corrections are included, particularly under individual attacks. 

\section{Conclusion}

The central result of this work is that the heralding protocol
substantially improves tolerance to channel fading under the individual
attacks considered. Compared with the squeezed-state homodyne and
coherent-state heterodyne benchmarks, the heralded protocol maintains
positive secret key rates over substantially greater channel losses
by suppressing the effect of channel fading. This improvement demonstrates that the
choice of state preparation and heralding conditions can strongly
influence the sensitivity of the key rate to transmission fluctuations.

This robustness is particularly relevant to free-space quantum communication, where fluctuating channel transmission can limit the
performance of otherwise favourable CVQKD protocols. The results suggest that CVQKD, secure against individual attacks, could be implemented via a satellite to ground downlink using our heralded scheme and motivate further investigation of heralded schemes for atmospheric
channels, including satellite-based links.  Under collective attacks, however, the secret key rate remains
strongly affected by fading. The difference between the individual- and collective-attack results identifies an important limitation of the present scheme: the favourable behaviour obtained within the individual-attack model does not extend to comparable robustness against collective attacks, although it is appropriate to note that it is unlikely that any adversary currently possesses the technology needed to perform collective attacks on CVQKD. A natural
direction for further work is to investigate whether alternative
heralding conditions or encoding strategies achieve fading channel noise suppression under collective attacks. This work represents a promising step towards satellite-based quantum communication by informing the development of CVQKD protocols that are more resilient to fading channels.

\section{Acknowledgements}
This research was supported by the Australian Research Council’s Linkage Projects funding scheme (Project No. LP200100601). J.D. acknowledges support from the Australian Government through an Australian Research Council Discovery Early Career Researcher Award, (Project No. DE250100914). We acknowledge helpful discussions with Matthew S. Winnel and Ryan J. Marshman. 

\begin{appendix}
	\section{Heralding protocol}
In this section, we provide calculations of this protocol. While the original scheme and calculations are contained in Ref.~\cite{winnel2021minimization}, for convenience, we reproduce the calculation here.  As shown in Fig.~\ref{fig:diagram}, the initial state of the system contains two two-mode-squeezed vacuum sources where modes $A1$ and $B$ are entangled and modes $A2$ and $A3$ are entangled. The covariance matrix of the four-mode state is given by:
\begin{equation}
	\nu \equiv \sqrt{\mu^2-1}.
\end{equation}
\begin{equation}
	\Gamma^{(0)}_{A_1A_2BA_3}
	=
	\begin{bmatrix}
		\mu\mathbf{I}_2 & \mathbf{0}_2 & \nu\mathbf{Z} & \mathbf{0}_2 \\
		\mathbf{0}_2 & \mu\mathbf{I}_2 & \mathbf{0}_2 & \nu\mathbf{Z} \\
		\nu\mathbf{Z} & \mathbf{0}_2 & \mu\mathbf{I}_2 & \mathbf{0}_2 \\
		\mathbf{0}_2 & \nu\mathbf{Z} & \mathbf{0}_2 & \mu\mathbf{I}_2
	\end{bmatrix},
\end{equation}
%\begin{equation}
%	\Gamma^{(0)}_{A_1A_2BA_3}
%	=
%	\begin{bmatrix}
%		\mu\mathbf{I}_2 & 0 & \nu\mathbf{Z} & 0 \\
%		0 & \mu\mathbf{I}_2 & 0 & \nu\mathbf{Z} \\
%		\nu\mathbf{Z} & 0 & \mu\mathbf{I}_2 & 0 \\
%		0 & \nu\mathbf{Z} & 0 & \mu\mathbf{I}_2
%	\end{bmatrix},
%\end{equation}
where $\nu=\sqrt{\mu^2-1}$ and 
\begin{equation}
	\mathbf{I}_2=\operatorname{diag}(1,1),
	\qquad
	\mathbf{Z}=\operatorname{diag}(1,-1),
	\qquad
	\mathbf{0}_2=\operatorname{diag}(0,0).
\end{equation}

Squeezing is then applied to modes $A3$ and $B$. The covariance matrix transforms as:
\begin{equation}
	\Gamma^{(1)}_{A_1A_2BA_3}
	=
	S\,\Gamma^{(0)}_{A_1A_2BA_3}\,S^{T},
\end{equation}
with 
\begin{equation}
	S
	=
	\begin{bmatrix}
		\mathbf{I}_2 & \mathbf{0}_2 & \mathbf{0}_2 & \mathbf{0}_2 \\
		\mathbf{0}_2 & \mathbf{I}_2 & \mathbf{0}_2 & \mathbf{0}_2 \\
		\mathbf{0}_2 & \mathbf{0}_2
		& \operatorname{diag}\!\left(e^{-r},e^{r}\right) & \mathbf{0}_2 \\
		\mathbf{0}_2 & \mathbf{0}_2 & \mathbf{0}_2
		& \operatorname{diag}\!\left(e^{r},e^{-r}\right)
	\end{bmatrix}.
\end{equation}
\begin{widetext}
\begin{equation}
	\Gamma^{(1)}_{A_1A_2BA_3}
	=
	\begin{bmatrix}
		\mu & 0 & 0 & 0 & \nu e^{-r} & 0 & 0 & 0 \\
		0 & \mu & 0 & 0 & 0 & -\nu e^{r} & 0 & 0 \\
		0 & 0 & \mu & 0 & 0 & 0 & \nu e^{r} & 0 \\
		0 & 0 & 0 & \mu & 0 & 0 & 0 & -\nu e^{-r} \\
		\nu e^{-r} & 0 & 0 & 0 & \mu e^{-2r} & 0 & 0 & 0 \\
		0 & -\nu e^{r} & 0 & 0 & 0 & \mu e^{2r} & 0 & 0 \\
		0 & 0 & \nu e^{r} & 0 & 0 & 0 & \mu e^{2r} & 0 \\
		0 & 0 & 0 & -\nu e^{-r} & 0 & 0 & 0 & \mu e^{-2r}
	\end{bmatrix}.
\end{equation}
%\end{widetext}
A 50:50 beamsplitter is then applied combining modes $A3$ and $B$. 
\begin{equation}
	\Gamma^{(2)}_{A_1A_2BA_3}
	=
	B\,\Gamma^{(1)}_{A_1A_2BA_3}\,B^{T},
\end{equation}
where
\begin{equation}
	B
	=
	\begin{bmatrix}
		1 & 0 & 0 & 0 & 0 & 0 & 0 & 0 \\
		0 & 1 & 0 & 0 & 0 & 0 & 0 & 0 \\
		0 & 0 & 1 & 0 & 0 & 0 & 0 & 0 \\
		0 & 0 & 0 & 1 & 0 & 0 & 0 & 0 \\
		0 & 0 & 0 & 0 & \frac{1}{\sqrt{2}} & 0
		& \frac{1}{\sqrt{2}} & 0 \\
		0 & 0 & 0 & 0 & 0 & \frac{1}{\sqrt{2}}
		& 0 & \frac{1}{\sqrt{2}} \\
		0 & 0 & 0 & 0 & -\frac{1}{\sqrt{2}} & 0
		& \frac{1}{\sqrt{2}} & 0 \\
		0 & 0 & 0 & 0 & 0 & -\frac{1}{\sqrt{2}}
		& 0 & \frac{1}{\sqrt{2}}
	\end{bmatrix}.
\end{equation}
%\begin{widetext}
\begin{equation}
	\Gamma^{(2)}_{A_1A_2BA_3}
	=
	\begin{bmatrix}
		\mu & 0 & 0 & 0
		& \frac{\nu e^{-r}}{\sqrt{2}} & 0
		& -\frac{\nu e^{-r}}{\sqrt{2}} & 0
		\\
		0 & \mu & 0 & 0
		& 0 & -\frac{\nu e^{r}}{\sqrt{2}}
		& 0 & \frac{\nu e^{r}}{\sqrt{2}}
		\\
		0 & 0 & \mu & 0
		& \frac{\nu e^{r}}{\sqrt{2}} & 0
		& \frac{\nu e^{r}}{\sqrt{2}} & 0
		\\
		0 & 0 & 0 & \mu
		& 0 & -\frac{\nu e^{-r}}{\sqrt{2}}
		& 0 & -\frac{\nu e^{-r}}{\sqrt{2}}
		\\
		\frac{\nu e^{-r}}{\sqrt{2}} & 0
		& \frac{\nu e^{r}}{\sqrt{2}} & 0
		& \mu\cosh(2r) & 0
		& \mu\sinh(2r) & 0
		\\
		0 & -\frac{\nu e^{r}}{\sqrt{2}}
		& 0 & -\frac{\nu e^{-r}}{\sqrt{2}}
		& 0 & \mu\cosh(2r)
		& 0 & -\mu\sinh(2r)
		\\
		-\frac{\nu e^{-r}}{\sqrt{2}} & 0
		& \frac{\nu e^{r}}{\sqrt{2}} & 0
		& \mu\sinh(2r) & 0
		& \mu\cosh(2r) & 0
		\\
		0 & \frac{\nu e^{r}}{\sqrt{2}}
		& 0 & -\frac{\nu e^{-r}}{\sqrt{2}}
		& 0 & -\mu\sinh(2r)
		& 0 & \mu\cosh(2r)
	\end{bmatrix}.
\end{equation}

An \(x\)-quadrature homodyne measurement is performed on mode \(A_3\).
The conditional covariance matrix is \cite{weedbrook2012gaussian}
\begin{equation}
	\Gamma^{(3)}_{A_1A_2B}
	=
	\Gamma^{(2)}_{A_1A_2B}
	-
	\sigma
	\left(
	\Pi_x \Gamma^{(2)}_{A_3}\Pi_x
	\right)^{\mathrm{MP}}
	\sigma^{T},
\end{equation}

where \(\Pi_x \) is the projector onto the \(x\) quadrature 
\begin{equation}
\Pi_x
=
\left[
\begin{array}{cc}
	1 & 0 \\
	0 & 0
\end{array}
\right],
\end{equation}
and MP denotes the Moore–Penrose pseudoinverse, and
\begin{equation}
	\Gamma^{(2)}_{A_1A_2BA_3}
	=
	\begin{bmatrix}
		\Gamma^{(2)}_{A_1A_2B} & \sigma \\
		\sigma^{T} & \Gamma^{(2)}_{A_3}
	\end{bmatrix}.
\end{equation}

The resulting covariance matrix for modes $A1$, $A2$ and $B$ before the channel  is~\cite{winnel2021minimization}:

%	\begin{widetext}
%		\begin{equation}
%			\Gamma_{A1 A2B|\mathrm{hom}. A3_x} = \begin{bmatrix}
%	\frac{e^4r \mu^2+1}{\mu \left(e^{4r}+1\right)} & 0  & \frac{e^{2r} \left(\mu^2-1\right)}{\mu \left(e^{4r}+1\right)}
%			\end{bmatrix}
%					\end{equation}
%		\end{widetext}
\begin{equation}
	\Gamma^{(3)}_{A_1A_2B}
	=
	\begin{bmatrix}
		\displaystyle
		\frac{\mu^{2}e^{4r}+1}
		{\mu\left(e^{4r}+1\right)}
		&
		0
		&
		\displaystyle
		\frac{\nu^{2}e^{2r}}
		{\mu\left(e^{4r}+1\right)}
		&
		0
		&
		\displaystyle
		\frac{\sqrt{2}\,\nu e^{3r}}
		{e^{4r}+1}
		&
		0
		\\[2ex]
		0
		&
		\mu
		&
		0
		&
		0
		&
		0
		&
		\displaystyle
		-\frac{\nu e^{r}}{\sqrt{2}}
		\\[2ex]
		\displaystyle
		\frac{\nu^{2}e^{2r}}
		{\mu\left(e^{4r}+1\right)}
		&
		0
		&
		\displaystyle
		\frac{\mu^{2}+e^{4r}}
		{\mu\left(e^{4r}+1\right)}
		&
		0
		&
		\displaystyle
		\frac{\sqrt{2}\,\nu e^{r}}
		{e^{4r}+1}
		&
		0
		\\[2ex]
		0
		&
		0
		&
		0
		&
		\mu
		&
		0
		&
		\displaystyle
		-\frac{\nu e^{-r}}{\sqrt{2}}
		\\[2ex]
		\displaystyle
		\frac{\sqrt{2}\,\nu e^{3r}}
		{e^{4r}+1}
		&
		0
		&
		\displaystyle
		\frac{\sqrt{2}\,\nu e^{r}}
		{e^{4r}+1}
		&
		0
		&
		\displaystyle
		\frac{2\mu e^{2r}}
		{e^{4r}+1}
		&
		0
		\\[2ex]
		0
		&
		\displaystyle
		-\frac{\nu e^{r}}{\sqrt{2}}
		&
		0
		&
		\displaystyle
		-\frac{\nu e^{-r}}{\sqrt{2}}
		&
		0
		&
		\mu\cosh(2r)
	\end{bmatrix}.
	\label{eq:gamma3}
\end{equation}
\end{widetext}

After homodyne detection is performed on mode $A3$, mode $B$ is sent through a free space link to Bob where
\begin{equation}
	\operatorname{Var}\!\left(\sqrt{\eta}\right)
	=
	\left\langle \eta\right\rangle
	-
	\left\langle \sqrt{\eta}\right\rangle^{2}.
\end{equation}

According to \eqref{eq:cmFad}-\eqref{eq:bpFad}, the fading channel transforms the covariance matrix \ref{eq:gamma3} as:
\begin{widetext}
\begin{equation}
	\Gamma_{A_1A_2B}^{\mathrm{fad}}
	=
	\begin{bmatrix}
		\displaystyle
		\frac{\mu^2e^{4r}+1}{\mu(e^{4r}+1)}
		& 0
		& \displaystyle\frac{\nu^2e^{2r}}{\mu(e^{4r}+1)}
		& 0
		& \displaystyle\frac{\sqrt{2T}\,\nu e^{3r}}{e^{4r}+1}
		& 0
		\\[2ex]
		0 & \mu & 0 & 0 & 0
		& \displaystyle-\frac{\sqrt{T}\,\nu e^r}{\sqrt{2}}
		\\[2ex]
		\displaystyle
		\frac{\nu^2e^{2r}}{\mu(e^{4r}+1)}
		& 0
		& \displaystyle\frac{\mu^2+e^{4r}}{\mu(e^{4r}+1)}
		& 0
		& \displaystyle\frac{\sqrt{2T}\,\nu e^r}{e^{4r}+1}
		& 0
		\\[2ex]
		0 & 0 & 0 & \mu & 0
		& \displaystyle-\frac{\sqrt{T}\,\nu e^{-r}}{\sqrt{2}}
		\\[2ex]
		\displaystyle
		\frac{\sqrt{2T}\,\nu e^{3r}}{e^{4r}+1}
		& 0
		& \displaystyle\frac{\sqrt{2T}\,\nu e^r}{e^{4r}+1}
		& 0 & b'_x & 0
		\\[2ex]
		0
		& \displaystyle-\frac{\sqrt{T}\,\nu e^r}{\sqrt{2}}
		& 0
		& \displaystyle-\frac{\sqrt{T}\,\nu e^{-r}}{\sqrt{2}}
		& 0 & b'_p
	\end{bmatrix}.
	\label{eq:A1A2Bfad}
\end{equation}
where
\begin{align}
	\label{eq:bxfad}
	b'_x
	&=
	T\frac{2\mu e^{2r}}{e^{4r}+1}
	+1-T
	+
	\operatorname{Var}\!\left(\sqrt{\eta}\right)
	\left(
	\frac{2\mu e^{2r}}{e^{4r}+1}-1
	\right)
	+T\xi,
	\\
	b'_p
	&=
	T\mu\cosh(2r)
	+1-T
	+
	\operatorname{Var}\!\left(\sqrt{\eta}\right)
	\left[
	\mu\cosh(2r)-1
	\right]
	+T\xi.
	\label{eq:bpfad}
\end{align}

From \eqref{eq:A1A2Bfad} and \eqref{eq:bxfad}, the variance $V_{B_x\mid A_{1,x}}$ remaining after conditioning on Alice's $x$-quadrature measurement of mode $A_1$ is:
\begin{equation}
V_{B_x\mid A_{1,x}}=
1
+
\operatorname{Var}\!\left(\sqrt{\eta}\right)
\left(
-1+\frac{2e^{2r}\mu}{1+e^{4r}}
\right)
+
\frac{
	T\left[
	-1+2e^{2r}\mu
	+e^{4r}\mu^2(-1+\xi)
	+\xi
	\right]
}{
	1+e^{4r}\mu^2
}.
\label{eq:VBA1x}
\end{equation}
The mutual information can then be calculated via \eqref{eq:IAB} where $V_{B_x}$ is given by \eqref{eq:bxfad} and $V_{B_x\mid A_{1,x}}$ is given in \eqref{eq:VBA1x}. 
\subsection{Collective attack}
The Holevo bound on Eve's information is calculated from the symplectic eigenvalues of the covariance matrices before and after Bob's measurement. The Holevo quantity is
\begin{equation}
	\chi_{EB}
	=
	S\left(\Gamma_{A_1A_2B}^{\mathrm{fad}}\right)
	-
	S\left(\Gamma_{A_1A_2\mid B_x}^{\mathrm{fad}}\right),
	\label{eq:chiEBApp}
\end{equation}
where $\Gamma_{A_1A_2B}^{\mathrm{fad}}$ is the covariance matrix following Alice's heralding measurement and transmission of mode $B$ through the fading channel \eqref{eq:A1A2Bfad}-\eqref{eq:bpfad}. The covariance matrix  $	\Gamma_{A_1A_2\mid B_x}^{\mathrm{fad}}$ describes Alice's remaining modes conditioned on Bob's measurement of the $x$ quadrature and is obtained from
\begin{equation}
	\Gamma_{A_1A_2\mid B_x}^{\mathrm{fad}}
	=
	\Gamma_{A_1A_2}^{\mathrm{fad}}
	-
	\sigma
	\left(
	\Pi_x
	\Gamma_{B}^{\mathrm{fad}}
	\Pi_x
	\right)^{-1}
	\sigma^{T},
\end{equation}
where the submatrix is defined:
\begin{equation}
	\begin{aligned}
		\Gamma_{A_1A_2B}^{\mathrm{fad}}
		&=
		\left[
		\begin{array}{cc}
			\Gamma_{A_1A_2}^{\mathrm{fad}} & \sigma \\
			\sigma^{T} & \Gamma_{B}^{\mathrm{fad}}
		\end{array}
		\right],
	\end{aligned}
\end{equation}
\end{widetext}
The first entropy in \eqref{eq:chiEBApp} is evaluated from the three symplectic eigenvalues of $\Gamma_{A_1A_2B}^{\mathrm{fad}}$, while the second is evaluated from the two symplectic eigenvalues of $\Gamma_{A_1A_2\mid B_x}^{\mathrm{fad}}$.

\subsection{Individual entangling-cloner attack \label{sec:IBE}}

To evaluate Eve's information under an individual Gaussian attack, we model the effective fading channel  using an entangling cloner \cite{usenko2022redundancy}. Eve prepares a two-mode squeezed-vacuum state in modes $E_1$ and $E_2$, mixes mode $E_1$ with the transmitted mode $B$ on a beam splitter of transmissivity $T$, and forwards the output mode $B'$ to Bob. Eve retains the modes $E_1'$ and $E_2$ and measures their $x$ quadratures separately for every channel use. The resulting classical data may then be combined to estimate Bob's homodyne outcome.

Before transmission, the covariance matrix shared by Alice and Bob is written in the quadrature ordering $(x_A,p_A,x_B,p_B)$ as
\begin{equation}
	\Gamma_{AB}
	=
	\begin{bmatrix}
		a_x & 0   & c_x & 0   \\
		0   & a_p & 0   & c_p \\
		c_x & 0   & b_x & 0   \\
		0   & c_p & 0   & b_p
	\end{bmatrix}.
\end{equation}
Eve's two-mode squeezed-vacuum state has covariance matrix
\begin{equation}
	\Gamma_{E_1E_2}
	=
	\begin{bmatrix}
		\zeta & 0 & \sqrt{\zeta^2-1} & 0 \\
		0 & \zeta & 0 & -\sqrt{\zeta^2-1} \\
		\sqrt{\zeta^2-1} & 0 & \zeta & 0 \\
		0 & -\sqrt{\zeta^2-1} & 0 & \zeta
	\end{bmatrix},
\end{equation}
where $\zeta$ is the variance of either of Eve's individual modes. Since Eve's state is initially independent of the state shared by Alice and Bob, the total input covariance matrix is
\begin{equation}
	\Gamma_{ABE_1E_2}^{\mathrm{in}}
	=
	\Gamma_{AB}\oplus\Gamma_{E_1E_2}.
\end{equation}
The beam-splitter interaction between modes $B$ and $E_1$ is described by
\begin{equation}
	\Gamma_{AB'E_1'E_2}
	=
	S_{BE_1}
	\Gamma_{ABE_1E_2}^{\mathrm{in}}
	S_{BE_1}^{T}.
\end{equation}

\begin{widetext}
	\begin{equation}
		\resizebox{0.98\textwidth}{!}{%
			$\displaystyle
			\Gamma_{AB'E_1'E_2}
			=
			\begin{bmatrix}
				a_x
				& 0
				& \sqrt{T}\,c_x
				& 0
				& -\sqrt{1-T}\,c_x
				& 0
				& 0
				& 0
				\\
				0
				& a_p
				& 0
				& \sqrt{T}\,c_p
				& 0
				& -\sqrt{1-T}\,c_p
				& 0
				& 0
				\\
				\sqrt{T}\,c_x
				& 0
				& Tb_x+(1-T)\zeta
				& 0
				& \sqrt{T(1-T)}(\zeta-b_x)
				& 0
				& \sqrt{(1-T)(\zeta^2-1)}
				& 0
				\\
				0
				& \sqrt{T}\,c_p
				& 0
				& Tb_p+(1-T)\zeta
				& 0
				& \sqrt{T(1-T)}(\zeta-b_p)
				& 0
				& -\sqrt{(1-T)(\zeta^2-1)}
				\\
				-\sqrt{1-T}\,c_x
				& 0
				& \sqrt{T(1-T)}(\zeta-b_x)
				& 0
				& (1-T)b_x+T\zeta
				& 0
				& \sqrt{T(\zeta^2-1)}
				& 0
				\\
				0
				& -\sqrt{1-T}\,c_p
				& 0
				& \sqrt{T(1-T)}(\zeta-b_p)
				& 0
				& (1-T)b_p+T\zeta
				& 0
				& -\sqrt{T(\zeta^2-1)}
				\\
				0
				& 0
				& \sqrt{(1-T)(\zeta^2-1)}
				& 0
				& \sqrt{T(\zeta^2-1)}
				& 0
				& \zeta
				& 0
				\\
				0
				& 0
				& 0
				& -\sqrt{(1-T)(\zeta^2-1)}
				& 0
				& -\sqrt{T(\zeta^2-1)}
				& 0
				& \zeta
			\end{bmatrix}
			$%
		}
	\end{equation}

Because Bob obtains the raw key by homodyning the $x$ quadrature, it is sufficient to consider the $x$-quadrature covariance matrix. In the ordering
$(x_A,x_{B'},x_{E_1'},x_{E_2})$, this matrix is

	\begin{equation}
		\Gamma_{AB'E_1'E_2}^{(x)}
		=
		\begin{bmatrix}
			a_x
			&
			\sqrt{T}\,c_x
			&
			-\sqrt{1-T}\,c_x
			&
			0
			\\
			\sqrt{T}\,c_x
			&
			Tb_x+(1-T)\zeta
			&
			\sqrt{T(1-T)}(\zeta-b_x)
			&
			\sqrt{(1-T)(\zeta^2-1)}
			\\
			-\sqrt{1-T}\,c_x
			&
			\sqrt{T(1-T)}(\zeta-b_x)
			&
			(1-T)b_x+T\zeta
			&
			\sqrt{T(\zeta^2-1)}
			\\
			0
			&
			\sqrt{(1-T)(\zeta^2-1)}
			&
			\sqrt{T(\zeta^2-1)}
			&
			\zeta
		\end{bmatrix}.
	\end{equation}
\end{widetext}

For the effective fading channel considered here,
\begin{equation}
	T=\left\langle\sqrt{\eta}\right\rangle^2,
	\qquad
	\operatorname{Var}(\sqrt{\eta})
	=
	\langle\eta\rangle
	-
	\left\langle\sqrt{\eta}\right\rangle^2,
\end{equation}
and Bob's output variance is
\begin{equation}
	b_x'
	=
	Tb_x+1-T+T\xi
	+
	\operatorname{Var}(\sqrt{\eta})(b_x-1).
\end{equation}
The entangling-cloner model gives
\begin{equation}
	b_x'=Tb_x+(1-T)\zeta.
\end{equation}
The variance of Eve's two-mode squeezed state is therefore selected to reproduce the complete effective noise measured in Bob's $x$ quadrature:
\begin{equation}
	\zeta
	=
	1+
	\frac{
		T\xi+\operatorname{Var}(\sqrt{\eta})(b_x-1)
	}{
		1-T
	}.
\end{equation}

The covariance matrix of Eve's measured $x$ quadratures is
\begin{equation}
	\Gamma_E^{(x)}
	=
	\begin{bmatrix}
		(1-T)b_x+T\zeta
		&
		\sqrt{T(\zeta^2-1)}
		\\
		\sqrt{T(\zeta^2-1)}
		&
		\zeta
	\end{bmatrix},
\end{equation}
while their correlations with Bob's measurement are collected in
\begin{equation}
	\boldsymbol{\sigma}_{B_xE}
	=
	\begin{bmatrix}
		\sqrt{T(1-T)}(\zeta-b_x)
		&
		\sqrt{(1-T)(\zeta^2-1)}
	\end{bmatrix}.
\end{equation}
Consequently, the variance of Bob's outcome conditioned on Eve's two homodyne results is obtained from
\begin{align}
	V_{B'_x\mid E'_{1,x},E_{2,x}}
	&=
	b_x'
	-
	\boldsymbol{\sigma}_{B_xE}
	\left(\Gamma_E^{(x)}\right)^{-1}
	\boldsymbol{\sigma}_{B_xE}^{T}
	\\
	&=
	\left[
	\frac{T}{b_x}+(1-T)\zeta
	\right]^{-1}.
\end{align}
Eve's information about Bob's $x$-quadrature outcome is therefore
\begin{align}
	I_{BE}^{\mathrm{ind}}
	&=
	\frac{1}{2}
	\log_2
	\left(
	\frac{V_{B'_x}}
	{V_{B'_x\mid E'_{1,x},E_{2,x}}}
	\right)
	\\
	&=
	\frac{1}{2}
	\log_2
	\left[
	\left(Tb_x+(1-T)\zeta\right)
	\left(
	\frac{T}{b_x}+(1-T)\zeta
	\right)
	\right].
\end{align}

This construction reproduces the first and second moments of the ensemble-averaged fading channel in the quadrature used to generate the key. Since the ensemble produced by fading is generally non-Gaussian, the resulting expression should be interpreted as Eve's information for the corresponding effective Gaussian entangling-cloner attack.

\subsection{Finite-size effects }
We evaluate finite-size effects using the method of
Ref.~\cite{jain2022practical}. The probability of successful
information reconciliation is
\begin{equation}
	p_{\mathrm{f}}=1-\mathrm{FER},
\end{equation}
where \(\mathrm{FER}\) is the frame-error rate. The number of
symbols remaining after successful reconciliation is therefore
\(p_{\mathrm{f}}N\).

Statistical uncertainty in parameter estimation is incorporated
through the estimates used to calculate Eve's information. For
collective attacks, we use the covariance-matrix confidence
bounds to construct the
parameter-estimation-adjusted covariance matrix \cite{jain2022practical} . The corresponding
Holevo information is calculated from the unconditional and
conditional covariance matrices. For the heralded protocol with
homodyne detection, this gives
\begin{equation}
	\chi_{EB}^{\mathrm{PE}}
	=
	S\!\left(\Gamma_{A_1A_2B}^{\mathrm{PE}}\right)
	-
	S\!\left(\Gamma_{A_1A_2|B_x}^{\mathrm{PE}}\right),
	\label{eq:holevo_pe}
\end{equation}
where \(S\) denotes the von Neumann entropy. Both covariance
matrices are obtained from the same parameter-estimation-adjusted
joint covariance matrix. For the comparison protocols, we use
the corresponding covariance matrices and Bob's appropriate
homodyne or heterodyne measurement.

The
asymptotic-equipartition correction used in the calculation is 
\begin{equation}
	\Delta_{\mathrm{AEP}}
	=
	4(d_{\mathrm{rx}}+1)
	\sqrt{
		\log_2\!\left[
		\frac{2}
		{\left(p_{\mathrm{f}}\varepsilon_s^2/3\right)^2}
		\right]
	},
\end{equation}
where \(d_{\mathrm{rx}}\) is the digitisation depth per quadrature
and \(\varepsilon_s\) is the smoothing parameter. The correction
associated with uncertainty in the empirical entropy is
\begin{equation}
	\Delta_{\mathrm{ent}}
	=
	\sqrt{
		2\log_2\!\left(
		\frac{2}{\varepsilon_{\mathrm{ent}}}
		\right)
	},
\end{equation}
where \(\epsilon_{\mathrm{ent}}\) is the entropy-estimation
failure probability.

Conditioning on successful information reconciliation contributes
\begin{equation}
	\Delta_S
	=
	\log_2\!\left(
	p_{\mathrm{f}}
	-
	\frac{p_{\mathrm{f}}\varepsilon_s^2}{3}
	\right),
\end{equation}
while the privacy-amplification contribution is
\begin{equation}
	\Delta_H
	=
	2\log_2\!\left(
	\sqrt{2}\,\varepsilon_h
	\right),
\end{equation}
with \(\varepsilon_h\) the hashing parameter. Parameters have been chosen following recent works \cite{jain2022practical, zaunders2025enhanced}. In our numerical calculations, we use a reconciliation efficiency
of \(\beta=0.95\), and a reconciliation success probability of
\(p_{\mathrm{ec}}=0.9964\). The digitisation depth is
\(d_{\mathrm{rx}}=6\) bits per quadrature, and the security
parameters are set to
\[
\varepsilon_{\mathrm{PE}}
=\varepsilon_{\mathrm{ent}}
=\varepsilon_s
=\varepsilon_h
=10^{-10}.
\]

\end{appendix}

\end{document}